\documentclass{main}

\shorttitle{Spin valve effect in two-dimensional VSe$_2$ system}
\shortauthors{Jafari et al.}
\graphicspath{{./}}
\begin{document}

\title{Spin valve effect in two-dimensional VSe$_2$ system}

\author[0000-0001-8320-652X]{M.A.Jafari}
\affiliation{Department of Mesoscopic Physics, ISQI, Faculty of Physics, Adam Mickiewicz University, ul. Uniwersytetu Poznańskiego 2, 61-614 Pozna\'n, Poland}

\author[0000-0002-0594-775X]{M. Wawrzyniak-Adamczewska}
\affiliation{Department of Mesoscopic Physics, ISQI, Faculty of Physics, Adam Mickiewicz University, ul. Uniwersytetu Poznańskiego 2, 61-614 Pozna\'n, Poland}

\author[0000-0003-1475-5079]{S. Stagraczy\'nski}
\affiliation{Department of Mesoscopic Physics, ISQI, Faculty of Physics, Adam Mickiewicz University, ul. Uniwersytetu Poznańskiego 2, 61-614 Pozna\'n, Poland}

\author[0000-0002-1452-7001]{A. Dyrdal}
\affiliation{Department of Mesoscopic Physics, ISQI, Faculty of Physics, Adam Mickiewicz University, ul. Uniwersytetu Poznańskiego 2, 61-614 Pozna\'n, Poland}

\author[0000-0001-7823-6388]{J. Barna\'s}
\affiliation{Department of Mesoscopic Physics, ISQI, Faculty of Physics, Adam Mickiewicz University, ul. Uniwersytetu Poznańskiego 2, 61-614 Pozna\'n, Poland}
\affiliation{Institute of Molecular Physics, Polish Academy of Sciences, 60-179 Poznań, Poland}



\begin{abstract}

Vanadium based  dichalcogenides, VSe$_2$, are two-dimensional materials in which magnetic Vanadium atoms are arranged in a hexagonal lattice and are coupled ferromagnetically within the plane. However, adjacent atomic planes are coupled antiferromagnetically. This provides new and interesting opportunities for application in spintronics and data storage and processing technologies. A spin valve  magnetoresistance may be achieved when magnetic moments of both atomic planes are driven to parallel alignment by an external magnetic field. The resistance change associated with the transition from antiparallel to parallel configuration is qualitatively similar to that observed in artificially layered metallic  magnetic structures. Detailed electronic structure of VSe$_2$ was obtained from DFT calculations.
%, and from the total energy difference we estimated the interlayer exchange coupling.
Then, the ballistic spin-valve magnetoresistance was determined within the Landauer formalism.
In addition, we also analyze thermal and thermoelectric properties. Both phases of VSe$_2$, denoted as H and T, are considered.

\end{abstract}

\keywords{Van der Waals materials --- dichalcogenides  --- magnetoresistance ---  spin valves}


\section{Introduction} \label{sec:introduction}
Spin-valve magnetoresistance is a phenomenon that has been observed first in antiferromagnetically exchange-coupled metallic Fe/Cr/Fe bilayers and superlattices~\cite{baibich1988giant,binasch1989enhanced}. In the simplest case, magnetic moments of the two magnetic Fe layers are oriented in opposite orientations (in the film plane), while an external in-plane magnetic field drives the moments to parallel configuration. The electrical  resistance of such a system decreases during rotation of the moments from antiparallel to parallel configurations. This resistance drop is known as the spin-valve magnetoresistance (or equivalently giant magnetoresistance)~\cite{baibich1988giant,binasch1989enhanced,barnas1990novel}. The phenomenon turned out to be very useful for practical applications as  sensors of weak magnetic fields in hard-disc drives~\cite{dieny1991giant}. Many various realisations of spin valves have been proposed in recent years, including vertically magnetized ultrathin layered metallic systems, tunnel junctions~\cite{Yuasa2004}, magnetic single-electron transistors~\cite{barnas1998magnetoresistance}, molecular junctions~\cite{zhu2010molecular,Krompiewski2004}, and others. In all cases there are two magnetic elements (usually electrodes), whose magnetic moments can be rotated in a controllable way from antiparallel to parallel configurations.

Further miniaturization of nanoelectronics and spintronics devices, especially in view of recent developments in the information storage and processing technologies, stimulates search for novel materials that can be used to construct new spin-vales. An interesting and promising solutions rely on novel two-dimensional crystals, especially on van der Waals magnetic materials, like for instance Vanadium Dichalcogenides VSe$_2$~\cite{Tong2017,li2014versatile,fuh2016metal,fuh2016newtype,Gong2018,jiang2020independent}.  The  key property of such  materials is a weak van-der-Waals  coupling between different atomic planes, so one can easily produce materials with well controlled number of atomic planes, down to single atomic monolayers~\cite{Burch2018,Aivazian2015,zhao2016colloidal,bayard1976anomalous}. A single monolayer of VSe$_2$ consists of one atomic plane of vanadium atoms  and two atomic sheets of Se atoms, that are shifted in opposite directions out of the atomic plane of V atoms. The Vanadium atoms in a single monolayer form a hexagonal lattice with ferromagnetic exchange coupling between nearest neighbours. Exchange coupling between next-nearest-neighbours is antiferromagnetic and is negligibly small. This leads to a strictly 2D ferromagnetic system. Adjacent monolayers of VSe$_2$ are coupled {\it via} van-der-Waals forces, and the V-atoms in different planes are  exchange-coupled antiferromagnetically. Thus, a bilayer of VSe$_2$ with antiferromagnetic interlayer coupling is conceptually similar to a metallic spin valve with two magnetic layers coupled through a nonmagnetic spacer due to antiferromagnetic indirect exchange coupling of RKKY-type. Such an interaction is antiferromagnetic in certain regions of the nonmagnetic spacer thicknesses.
It is clear now, that bilayer of VSe$_2$ is a natural atomically thin spin valve. In the ground state magnetic moments of both V-planes are oriented in opposite orientations perpendicular  to the planes. Relatively weak external magnetic field normal to the planes drives the magnetic moments of both atomic planes to parallel orientations, exactly like in conventional metallic spin valves. In this paper we show that this rotation of magnetic moments from antiparallel to parallel configuration is associated with a resistance change, i.e. there is a spin-valve magnetoresistance like in metallic systems. In general, the spin-valve magnetoresistance can be defined for current flowing in plane of the system (so-called CIP geometry) as well as when it flows perpendicularly to the planes (CPP geometry). In this paper we restrict considerations to the case when current flows in the 2D planes, i.e., we consider the CIP geometry.

To do this we calculate first the electronic structure of VSe$_2$ bilayer. Then, from the total energy calculations we confirm that  magnetic moments of the two V atomic planes are arranged antiferromagnetically  in the ground state (i.e. the interlayer exchange coupling is antiferromagnetic), in agreement with literature~\cite{Gong2018,li2020coupling,feng2018strain}. Since  VSe$_2$ bilayer exists in two different staking configurations, so-called H and T phases, we consider both of them. In the H phase the Vanadium atoms in different planes are one above the other~\cite{Esters2017}, while in the T phase they are displaced and each V atom has three nearest neighbours in the adjacent Vanadium plane~\cite{li2014versatile}. For both phases we  calculate transport characteristics in the ballistic transport regime. To find conductance, we calculate transmission coefficient and then  use the Landauer~\cite{Landauer1970} formula to find conductance.  From the conductance in both parallel (ferromagnetic) and antiparallel (antiferromagnetic) configurations we determine  magnitude of the magnetoresistance as a function of the position of Fermi level. In practice, this position can be generally controlled externally (e.g. by a gate voltage). We also calculate electronic contribution to the heat conductance and the corresponding thermal magnetoresistance. In addition, we also determine the thermoelectric efficiency~\cite{hatami2009thermoelectric,gravier2006thermodynamic}. More details on numerical method are in section \hyperref[sec:methodology]{2}. Numerical results on electronic band structure and transport characteristics (magnetoresistance) are given in sections \hyperref[sec:BS]{3} and \hyperref[sec:transport]{4}, respectively. In turn, final conclusions are in section \hyperref[sec:summary]{5}.

\section{Numerical method} \label{sec:methodology}

For transport analysis we take a nanoribbon of VSe$_2$ bilayer with a finite width, and assume  that the central part of the nanoribbon represents our system, while the left and right parts of the nanoribbon play the role of electrodes. To find transport properties we calculate first electronic structure of the nanoribbon, from which we determine the transmission function and subsequently all transport characteristics.
In this paper, the electronic structure -- and thus also transport properties -- are determined from first principle calculations within the framework of Density Functional Theory (DFT)~\cite{taylor2001ab}  and  nonequilibrium Green's functions (NEGFs) as implemented in  the  Quantum ATK-LCAO code package \cite{smidstrup2017first}. Within this approach, the generalized gradient approximation of Perdew-Burke Ernzerhof \cite{perdew1996generalized} is used to describe the exchange-correlation functional for the electrons. Moreover, an appropriate semi-empirical modelling correction known as Grimme DFT-D2 approximation~\cite{Grimme} was used to deal with the weak interlayer van der Waals interaction. The PseudoDojo collection of optimized norm-conserving Vanderbilt (ONCV) pseudopotentials and ultra basis set were used to optimize the structures and study the electronic properties~\cite{Setten}.
The Brillouin zone sampling was adopted using $12\times 12\times 1$
and $1\times 1\times 300$ Monkhorst-Pack grid for the electronic and transport calculations, respectively, and the mesh cut-off energy was set to 600 Ry. Furthermore, the force per atom less than ${10}^{- 3}$~eV/$\AA$
was minimized through the standard Conjugate-Gradients (CG) method. In addition, relative convergence for the Self-Consistent Field (SCF) energy is reached until
$10^{-5}$~eV/$\AA$.

We limit further analysis of transport characteristics to the  ballistic transport and linear response regime.
Using methods implemented in Quantum ATK + NEGF package~\cite{palsgaard2018efficient}, one finds the transmission function $\mathcal{T}(\varepsilon)$,
$\mathcal{T}(\varepsilon) = \sum_{\sigma} \mathrm{Tr}\left\{\Gamma_{L,\sigma}(\varepsilon)\mathcal{G}_{\sigma}^{R}(\varepsilon)\Gamma_{\sigma}^{R}(\varepsilon) \mathcal{G}_{\sigma}^{A}(\varepsilon)\right\} $.
Here $\mathcal{G}_{\sigma}^{R/A}(\varepsilon)$ denotes retarded and advanced Green's functions of the central region, $\Gamma_{L/R,\sigma}(\varepsilon)$ is the line-width broadening matrix due to spin-dependent coupling between the central scattering and left/right electrode respectively, and $\sigma$ is the spin index.
According to the Landauer-Buttiker formula,  the electric current in nonequilibrium situation is given as:
$I  = \frac{e}{h} \int d\varepsilon \mathcal{T}(\varepsilon) \left[f_{L}(\varepsilon ,  \mu, T) - f_{R}(\varepsilon ,  \mu, T)\right]$,
where $f_{L/R}(\varepsilon ,  \mu, T)$ is the Fermi-Dirac distribution function for  the left/right electrode respectively ($\mu$ and $T$ are  the corresponding chemical potential and  temperature). In equilibrium case (linear response), temperatures and chemical potentials of both electrodes are equal, $T_{L} = T_{R} = T$ and $\mu_{L} = \mu_{R} = \mu$, respectively, and transport parameters are determined by the
kinetic coefficients $L_n$ defined as follows:
\begin{equation}
L_n(\mu, T) = - \frac{1}{h}\int d\varepsilon \mathcal{T}(\varepsilon) (\varepsilon - \mu)^{n} \frac{\partial f(\varepsilon, \mu, T)}{\partial \varepsilon}
\end{equation}
for $n=0,1,2$.
Accordingly, the linear electrical conductance $G$ can be calculated as $G =e^2 L_{0}$, while  the electronic contribution to the thermal conductance, $\kappa_{el}$, and Seebeck coefficient (thermopower), $S$, are then given by the formulas:~\cite{uchida2013longitudinal,liu2014spin,liu2012proposal}:
\begin{equation}
\kappa_{el} =  \frac{1}{T} \left[ L_{2} - \frac{L_{1}^2}{L_0}\right],
\end{equation}
\begin{equation}
S = - \frac{1}{eT} \frac{L_{1}}{L_{0}}.
\end{equation}
As a result, the corresponding figure of merit can be written in the following form:
\begin{equation}
ZT = \frac{S^{2} G}{\kappa_{el}} T.
\end{equation}
In general, the denominator in the above equation for $ZT$ should also include the phonon contribution. As our main interest was in spin valves, the figure of merit was calculated only with the electronic contribution to the heat conductivity, just to evaluate the upper limit of ZT. Any phonon contribution  reduces the magnitude of $ZT$, though at low temperatures the phonon contribution is small and the electronic term dominates, so Eq.~4 describes reasonably well the parameter $ZT$ at low temperatures. At higher temperatures, however, one should include the phonon term in realistic calculations.

%%%%%%%%%%%%%%%%%%%%%%%%%%%%%%%%%%%%%%%

\begin{figure}[ht!]
\plotone{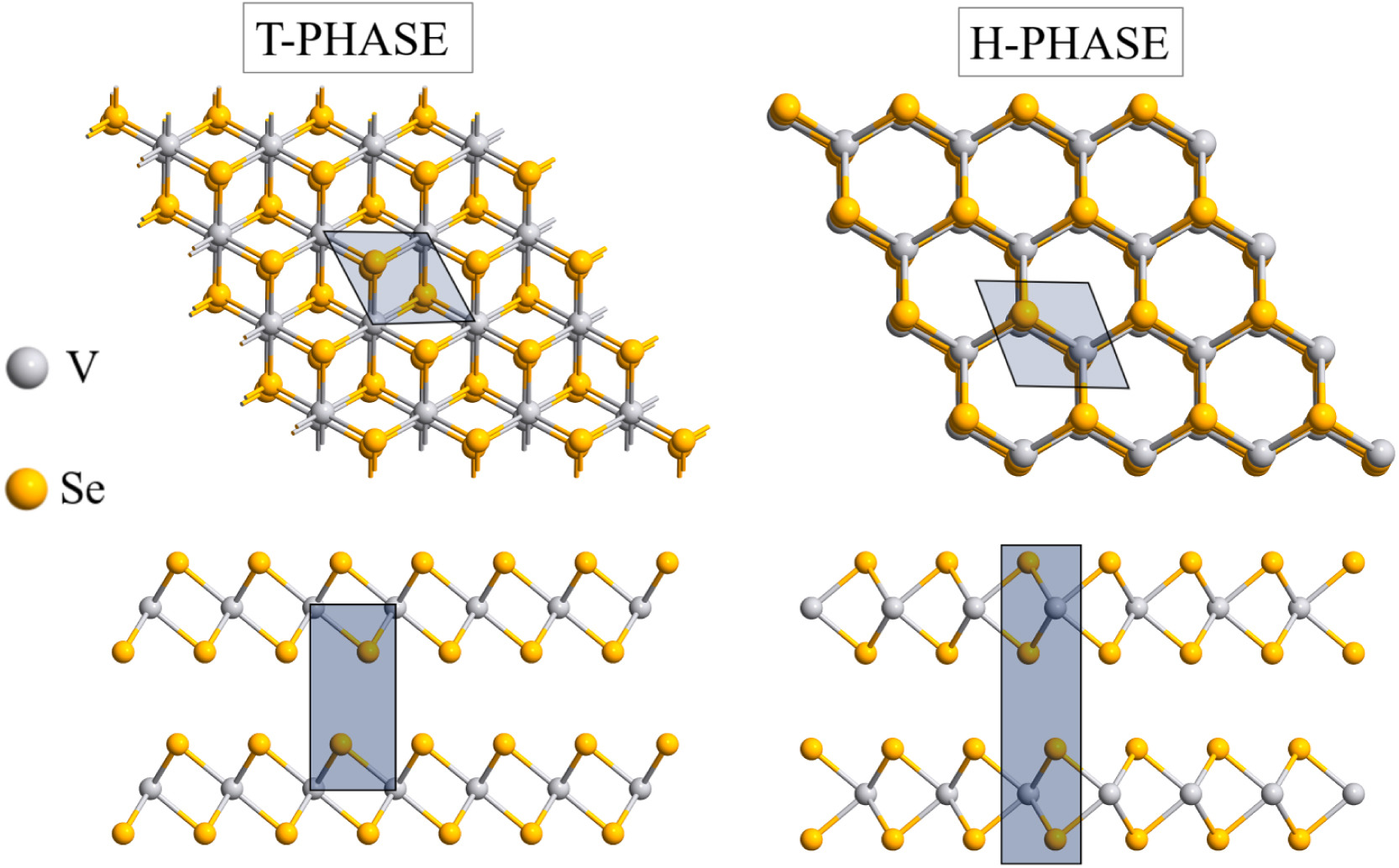}
\caption{Atomic structure of the 2D  VSe$_2$ bilayer in the T (left side) and H (right side) crystallographic phases. The upper parts show slightly tilted top views, while the bottom parts show the corresponding side views. Primitive unit cells are also marked by the shaded areas. In the T-phase bilayer the Vanadium atoms are one above the other (AA stacking), while in the H-phase bilayer the stacking is of AB type.  \label{fig:astr}}
\end{figure}

\section{Electronic band structure: 2D system} \label{sec:BS}

\begin{figure}[ht!]
\plotone{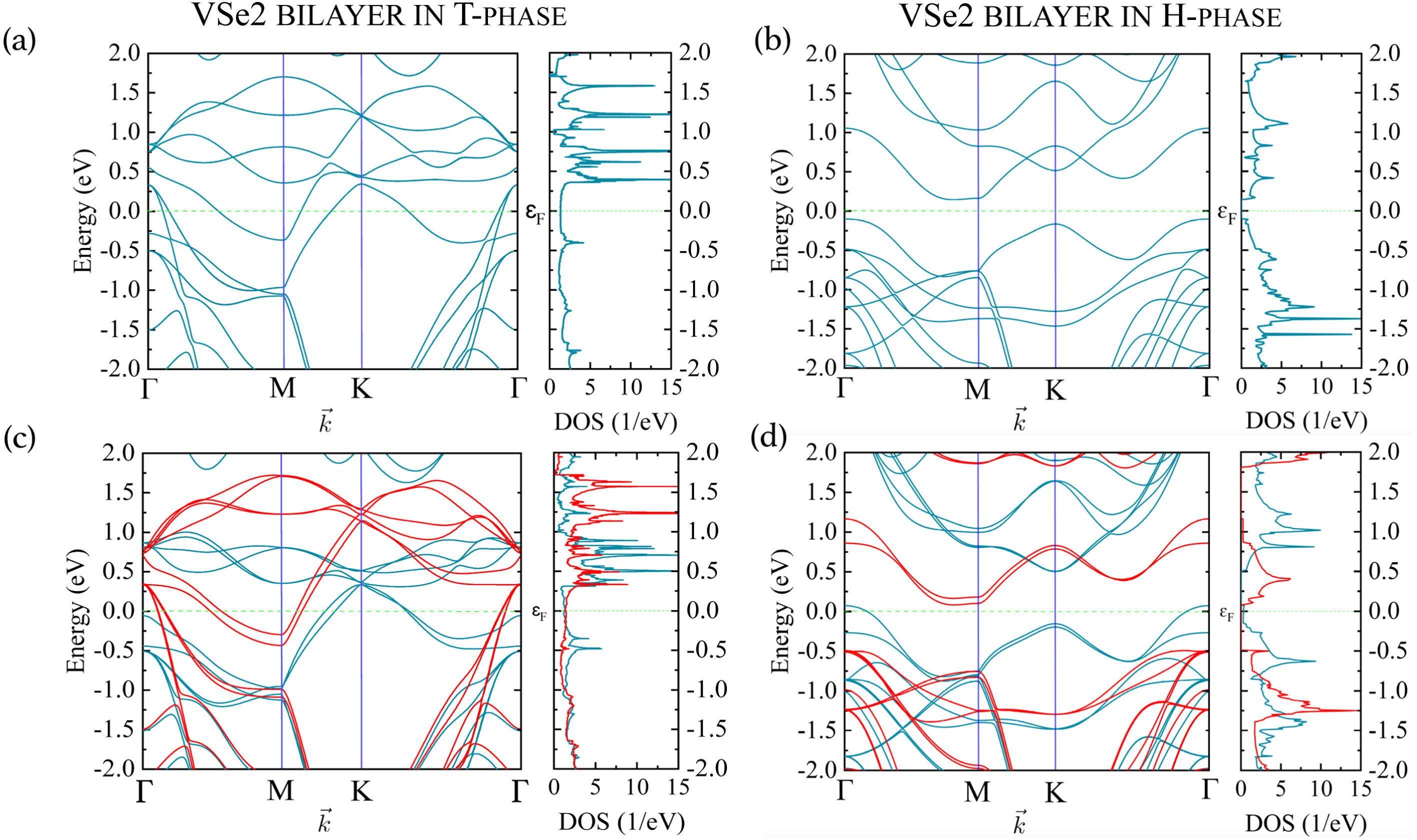}
\caption{Electronic structure of the 2D  VSe$_2$ bilayer in the T (a,c) and H (b,d) crystallographic phases  for the antiferromagnetic (a,b) and  ferromagnetic (c,d)  configurations of the magnetic moments of both vanadium atomic planes, and for the Coulomb parameter $U=0$. The electronic structure in the ferromagnetic state is spin dependent, while that in the antiferromagnetic state is spin degenerate. The corresponding density of states for both magnetic configurations is also shown on the right side. In the H phase, the ferromagnetic state is metallic, while a narrow gap appears in the antiferromagnetic state. In the T phase, both states are metallic.  \label{fig:fig1}}
\end{figure}

The bulk form of VSe$_2$ exists in two polymorphs, namely in the trigonal prismatic structure (2H, D3h) and in the octahedral (1T, D3d) structure. Atomic configuration of the T-phase and H-phase bilayers of VSe$_2$ is shown in Fig.\ref{fig:astr}. Note, in the case of T bilayers the stacking is of AA type, while for H bilayers it is of AB type. Accordingly, in the T-phase bilayer the Vanadium atoms are one above the other, while the H-phase bilayer consists of two hexagonal atomic layers  with an AB stacking, in which the V (Se) atoms of the top layer cover the Se (V) atoms of the bottom layer ~\cite{manzeli20172d,wasey2015quantum}. This can be clearly seen in the side views of the bilayers shown in the bottom parts of Fig.\ref{fig:astr}.

Based on available  literature, we assume a sufficiently strong exchange coupling of ferromagnetic type within atomic planes of Vanadium atoms \cite{li2020coupling,feng2018strain,Esters2017}, and focus only on the exchange coupling between adjacent atomic planes of VSe$_2$. To verify equilibrium magnetic configuration we calculate the electronic structure and total energy per unit cell in the parallel configuration (ferromagnetic state) and antiparallel one (antiferromagnetic state).
The spin-resolved electronic structure of the 2D VSe$_2$ bilayer in the T and H configurations is shown in Fig.\ref{fig:fig1} for both ferromagnetic and antiferromagnetic states. In the T phase, both antiferromagnetic (Fig.\ref{fig:fig1}(a)) and ferromagnetic (Fig.\ref{fig:fig1}(c)) states are metallic. In turn,  the ferromagnetic state in the H phase is  metallic (Fig.\ref{fig:fig1}(d)), while there is a narrow indirect gap in the antiferromagnetic state (Fig.\ref{fig:fig1}(b)).  The corresponding density of states per unit energy is also shown in Fig.\ref{fig:fig1} for both magnetic states and both T and H phases. This
density in the H phase clearly reveals a gap at the Fermi level of a neutral system
in the antiferromagnetic state.

From the total energy calculations one can evaluate interlayer exchange coupling and determine magnetic configuration in the ground state. In both T and H phases, the total energy in the ferommagnetic state is lager by than that  in the antiferromagnetic state, thus the ground state in both phases is antiferromagnetic. The corresponding antiferromagnetic exchange coupling parameter is pretty small, of an order of a few meV~\cite{Gong2018}.
Accordingly, the antiferromagnetic state becomes destabilized at relatively low temperatures. For practical realizations one would need materials with antiferromagnetic state stable at higher temperatures. This can be forced externally, for instance by exchange coupling to adjacent insulating magnetic layers with different coercive fields, so one can reach antiparallel and parallel configurations at higher temperatures and in appropriate external magnetic fields.

\section{Electron transport in a nanoribbon: spin-valve magnetoresistance} \label{sec:transport}

\begin{figure}[ht!]
\plotone{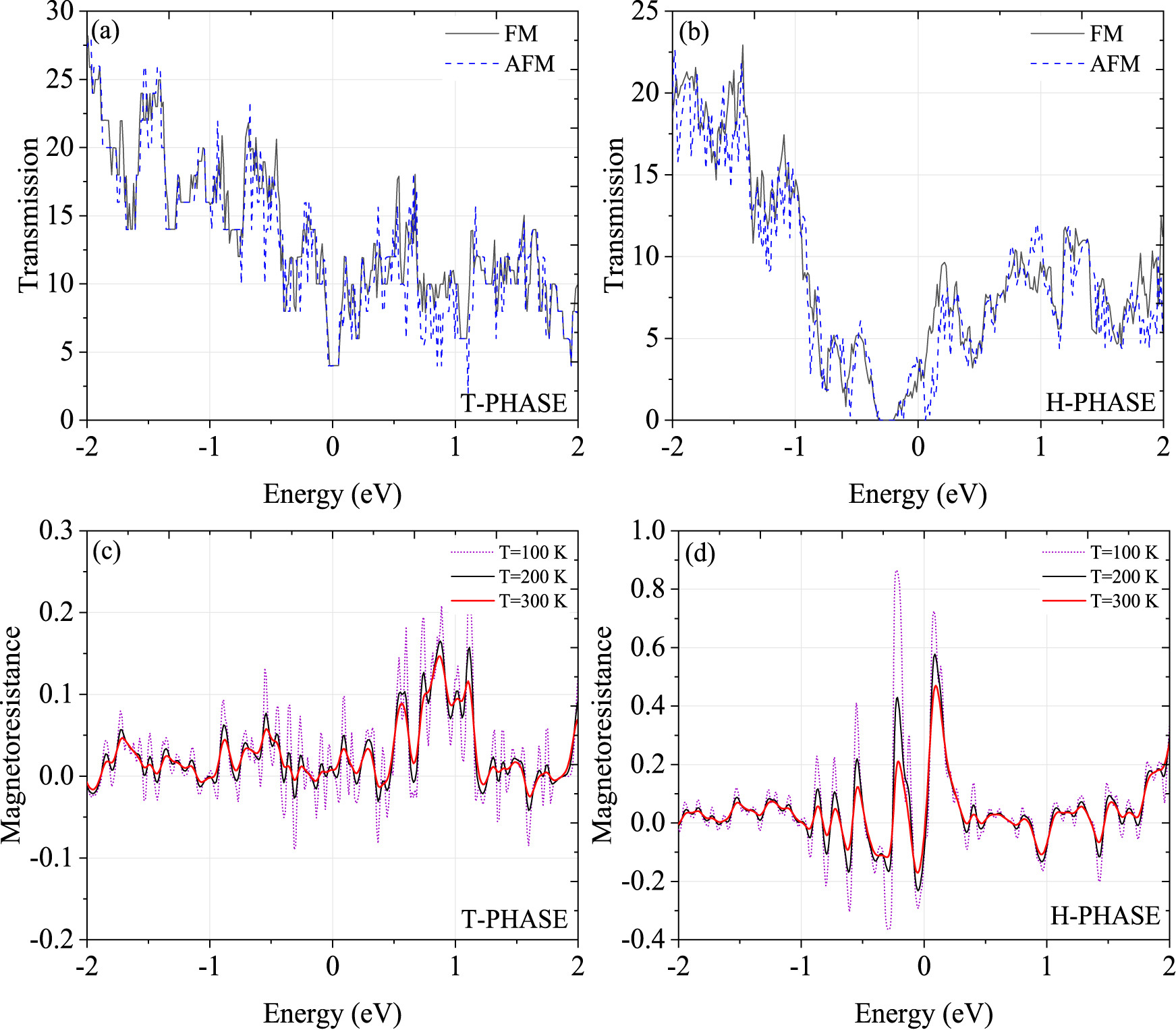}
\caption{Total transmission coefficient in the ferromagnetic and antiferromagnetic states for the T (a) and H (b) phases,  and the corresponding spin-valve magnetoresistance, (c) and (d), as a function of the energy $E$ for indicated temperatures. The energy $E$ can be considered as an externally tuneable Fermi energy (or chemical potential).  Fermi energy of a neutral (pristine) material corresponds to $E=0$.  \label{fig:fig2}}
\end{figure}

In the linear  response regime the conductance and  ballistic magnetoresistance are determined by the transmission coefficient at the Fermi energy. Generally, the Fermi energy (chemical potential) can be tuned by a gate voltage. In Fig.\ref{fig:fig2}(a,b)  we show the transmission coefficient for a nanoribbon in both ferromagnetic and antiferromagnetic states and for the T (a) and H (b) phases. The corresponding  linear  ballistic magnetoresistance as a function of the energy $E$  for both T and H phases is shown in Fig.\ref{fig:fig2}(c,d). Note, $E$ can be considered as a chemical potential or Fermi energy, and the Fermi energy of a pristine (neutral) material corresponds to $E=0$.
The magnetoresistance is defined quantitatively as the ratio of resistance difference in the two magnetic states normalized to the sum of both resistances,  ${\rm MR}=
(R_{\rm AP}-R_{\rm P})/(R_{\rm AP}+R_{\rm P})=
(G_{\rm P}-G_{\rm AP})/(G_{\rm P}+G_{\rm AP})$, where $R_{\rm AP}(R_{\rm P})$ stands for the  resistance in the antiparallel (parallel) state, while $G_{\rm AP}(G_{\rm P})$ is the corresponding conductance. The magnitude of magnetoresistance is relatively small at the Fermi level of neutral (pristine) systems, up to 10 percent. The magnitude of magnetoresistance varies with position of $E$ (chemical potential)  taking at maximum about $20\%$ in the T phase and $60\%$ in the H phase. Moreover, it acquires  positive as well as negative values. The magnetoresistance is shown there for three different temperatures, and as  one can note, the increasing temperature smooths the dependence on the energy.

\begin{figure}[ht!]
\plotone{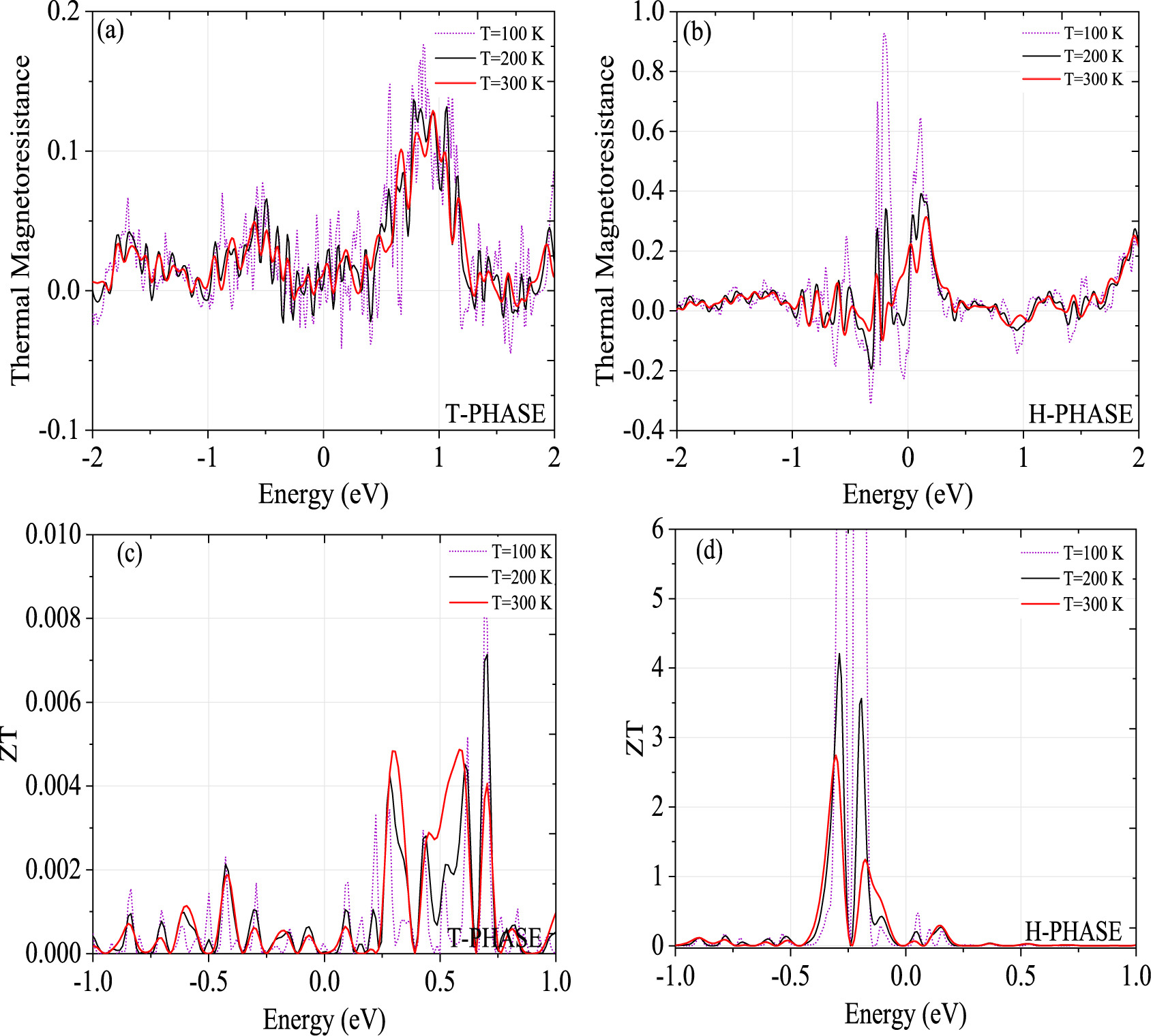}
\caption{Thermal magnetoresistance  as a function of  energy in the T (a) and H (b) phases. The corresponding thermoelectric efficiency (figure of merit)  $ZT$ is shown in (c) and (d).  \label{fig:fig3}}
\end{figure}

Similarly to the spin-valve magnetoresistance phenomenon described above, one may also define the thermal magnetoresistance (TMR) effect~\cite{tsyplyatyev2006thermally}, which reflects the fact that also electronic contribution to the heat conductance depends on the magnetic state, and thus the heat conductance changes with rotation from antiferromagnetic state to the ferromagnetic one. This thermal magnetoresistance may be defined in a similar way as the electrical one,
${\rm TMR}=
(\kappa_{\rm P}-\kappa_{\rm AP})/(\kappa_{\rm P}+\kappa_{\rm AP})$, where $\kappa_{\rm AP}$ and $\kappa_{\rm P}$ are the corresponding electronic contributions to the heat conductances in the antiparallel and parallel configurations. The thermal magnetoresistance in the linear response is shown in Fig.\ref{fig:fig3}(a,b) for both T and H phases. Similarly as MR, the magnitude of thermal magnetoresistance depends on the energy $E$ (chemical potential, Fermi level), taking  positive as well as negative values. The thermal magnetoresistance is shown there for three different temperatures, and  one can note that the increasing temperature smooths the dependence on the Fermi energy, similarly as in the case of MR.

Temperature gradient applied to the system generates a charge current. This phenomenon is known as the Seebeck effect~\cite{goldsmid2010introduction}, and is of high  interest  for thermoelectric applications~\cite{dubi2009thermospin,boona2014spin,uchida2016thermoelectric}. In the linear response, the Seebeck coefficient (thermopower) is given by Eq.(3).  The corresponding thermoelectric efficiency $ZT$ (figure of merit) is shown for both phases in Fig.\ref{fig:fig3}(c,d). As one can note, $ZT$ is very small for the T phase but is remarkably  higher for the H phase. Thus, only the H phase may be of some interest for thermoelectric applications.  Since the current in ferromagnetic states is spin polarized due to spin-dependent electronic structure, one may also define spin thermoelectric phenomena~\cite{uchida2016thermoelectric,Swirkowicz2009,Zberecki2013}. This problem, however, is not considered in this paper.

\section{Summary and conclusions} \label{sec:summary}

We have analyzed numerically magnetoresistance in spin-valves based on 2D van der Waals magnetic materials. As an example we have taken a bilayer of Vanadium dichalcogenide VSe$_2$. Each atomic monolayer  of the VSe$_2$ bilayer is ferromagnetic, but different monolayers are coupled antiferromagnetically. Thus, the equilibrium configuration is antiferromagnetic, and external magnetic field can rotate  magnetic moments of the two monolayers from antiparallel to parallel state. This rotation is accompanied by a resistance change for in-plane transport. Such a behavior is qualitatively similar to that in metallic spin valves. Similar resistance change can be also observed for current flowing perpendicularly to the atomic planes. We have also analyzed the corresponding thermal magnetoresistance and thermoelectric phenomenon. The latter effect is rather small in the T phase, while it can be remarkable in the H phase, especially when tuning the Fermi level below that of pristine material.

We also note, that such free standing spin valves can work at low temperatures.  This is because magnetism of these materials is not stable at higher temperatures. To reach any real application, one needs materials with magnetism at higher temperatures (with stronger ferromagnetic exchange coupling within the planes and larger antiferromagnetic exchange coupling between the atomic planes).  Fortunately, owing to recent progress in synthesis of novel 2D materials, one may believe that these expectations will be fulfilled in new materials.

\break
\section{Acknowledgments} \label{sec:ref}
This work has been supported by the Norwegian Financial Mechanism 2014-
2021 under the Polish-Norwegian Research Project NCN GRIEG “2Dtronics”
no. 2019/34/H/ST3/00515.

\break

\bibliography{Ref}{}
\bibliographystyle{aasjournal}



\end{document}